\documentclass[pra,showpacs,floatfix,twocolumn,letterpaper,footinbib]{revtex4}
\usepackage{bm,amsmath}
\usepackage{dcolumn}
\usepackage{natbib}

\newcommand{\Eref}[1]{Eq.~(\ref{#1})}
\newcommand{\tref}[1]{Table~\ref{#1}}

\newcommand{\kms}{km~s$^{-1}\,$}
\newcommand{\daa}{$\Delta\alpha/\alpha$\,}
\newcommand{\dmm}{$\Delta\mu/\mu$\,}
\newcommand{\ms}{m~s$^{-1}\,$}

\begin{document}
\title{Mid- and far-infrared fine-structure line sensitivities to
hypothetical variability of the fine-structure constant}
 \author{M. G. Kozlov$^{1}$}
 \author{S. G. Porsev$^{1}$}
 \author{S. A. Levshakov$^{2}$}
 \author{D. Reimers$^{3}$}
 \author{P. Molaro$^{4}$}
 \affiliation{$^1$Petersburg Nuclear Physics Institute, Gatchina,
              188300, Russia}
 \affiliation{$^2$Ioffe Physico-Technical Institute, Politekhnicheskaya Str. 26,
              194021 St. Petersburg, Russia}
 \affiliation{$^3$Hamburger Sternwarte, Universit\"{a}t Hamburg, Gojenbergsweg 112,
              D-21029 Hamburg, Germany}
 \affiliation{$^4$Osservatorio Astronomico di Trieste, Instituto Nazionale
di Astrofisica, Via G. B. Tiepolo 11, 34131 Trieste, Italy}
\date{ \today }

\begin{abstract}
Sensitivity coefficients to temporal variation of the fine-structure
constant $\alpha$ for transitions between the fine-structure (FS)
sub-levels of the ground states of C~\textsc{i}, Si~\textsc{i},
S~\textsc{i}, Ti~\textsc{i}, Fe~\textsc{i}, N~\textsc{ii},
Fe~\textsc{ii}, O~\textsc{iii}, S~\textsc{iii}, Ar~\textsc{iii},
Fe~\textsc{iii}, Mg~\textsc{v}, Ca~\textsc{v}, Na~\textsc{vi},
Fe~\textsc{vi}, Mg~\textsc{vii}, Si~\textsc{vii}, Ca~\textsc{vii},
Fe~\textsc{vii}, and Si~\textsc{ix} are calculated. These
transitions lie in the mid- and far-infrared regions and can be
observed in spectra of high-redshift quasars and infrared bright
galaxies with active galactic nuclei. Using FS transitions to study
$\alpha$-variation over cosmological timescale allows to improve the
limit on \daa\ by several times as compared to contemporaneous
optical observations ($|\Delta\alpha/\alpha| < 10^{-5}$), and to
suppress considerably systematic errors of the radial velocity
measurements caused by the Doppler noise. Moreover, the far infrared
lines can be observed at redshifts $z \gtrsim 10$, far beyond the
range accessible to optical observations ($z\lesssim 4$). We have
derived a simple analytical expression which relates the FS
intervals and the sensitivity of the FS transitions to the change of
$\alpha$.
\end{abstract}
\maketitle

\section{Introduction}
\label{sect-1}

The problem of variability of fundamental physical constants has a
long history starting 70 years ago with publications by
\citet{Mil35} and \citet{Dir37}. The review of its current status is
given in \cite{Uza03,GIK07}. Recent achievements in laboratory
studies of the time-variation of fundamental constants are
described, for example, in Refs.~\cite{Lea07,FK07c}.

The variability of the dimensionless physical constants is usually
considered in the framework of the theories of fundamental
interactions such as string and M theories, Kaluza-Klein theories,
quintessence theories, etc. In turn, the experimental physics and
observational astrophysics offer possibilities to probe directly the
temporal changes in the physical constants both locally and at early
cosmological epochs comparable with the total age of the Universe
($T_{\rm U} = 13.8$ Gyr for the $H_0 = 70$ \kms Mpc$^{-1}$,
$\Omega_m = 0.3$, $\Omega_\Lambda = 0.7$ cosmology). Here we discuss
a possibility of using the ground state fine-structure (FS)
transitions in atoms and ions to probe the variability of
$\alpha$ at high redshifts, up to $z \sim 10$ ($\sim96$\% of $T_{\rm
U}$).

The constants which can be probed from astronomical spectra are the
proton-to-electron mass ratio, $\mu = m_{\rm p}/m_{\rm e}$, the
fine-structure constant, $\alpha = e^2/(\hbar c)$, or different
combinations of $\mu$, $\alpha$, and the proton gyromagnetic ratio
$g_{\rm p}$. The reported in the literature data concerning the
relative values of \dmm\ and \daa\ at $z\sim$~1--3 are controversial
at the level of a few ppm (1ppm = $10^{-6}$): \dmm = $24\pm6$ ppm
\cite{RBH06} versus $0.6\pm1.9$ ppm \cite{FK07a}, and
\daa = $-5.7\pm1.1$ ppm \cite{MWF03} versus $-0.6\pm0.6$ ppm \cite{SCP04},
$-0.4\pm1.9$ ppm \cite{QRL04}, and $5.4\pm2.5$ ppm \cite{LML07}.
Such a spread points unambiguously to the presence of unaccounted
systematics. Some of the possible problems were studied in
\cite{MRA07,MLM07,MWF07,SCP07}, but the revealed systematic errors
cannot explain the full range of the observed discrepancies between
the \daa\ and \dmm\ values. We can state, however, that a
conservative upper limit on the hypothetical variability of these
constants is $10^{-5}$.

Astronomical estimates of the dimensionless physical constants are
based on the comparison of the line centers in the
absorption/emission spectra of astronomical objects and the
corresponding laboratory values. In practice, in order to
disentangle the line shifts caused by the motion of the object and
by the putative effect of the variability of constants, lines with
different sensitivities to the constant variations should be
employed. However, if different elements are involved in the
analysis, an additional source of errors due to the so-called
Doppler noise arises. The Doppler noise is caused by non-identical
spatial distributions of different species. It introduces offsets
which can either mimic or obliterate a real signal. The evaluation
of the Doppler noise is a serious problem
\cite{Lev94,L04,BSS04,Car00,KCL05,LRK07}. For this reason lines
of a single element arising exactly from the same atomic or
molecular level are desired. This would provide reliable
astronomical constraints on variations of physical constants.

In the present communication we propose to use the mid- and
far-infrared FS transitions within the ground multiplets $^3\!P_J$,
$^5\!D_J$, $^6\!D_J$, $^3\!F_J$ and $^4\!F_J$ of some of the most
abundant atoms and ions, such as Si~\textsc{i}, S~\textsc{i},
Ti~\textsc{i}, Fe~\textsc{i}, Fe~\textsc{ii}, S~\textsc{iii},
Ar~\textsc{iii}, Fe~\textsc{iii}, Mg~\textsc{v}, Ca~\textsc{v},
Na~\textsc{vi}, Fe~\textsc{vi}, Mg~\textsc{vii}, Si~\textsc{vii},
Ca~\textsc{vii}, Fe~\textsc{vii}, and Si~\textsc{ix} for
constraining the variability of $\alpha$. This approach has the
following advantages. Most important is that each element provides
two, or more FS lines which can be used independently~--- this
considerably reduces the Doppler noise. The mid- and far-infrared FS
transitions are typically more sensitive to the change of $\alpha$
than optical lines. For high redshifts ($z > 2$), the far-infrared
(FIR) lines are shifted into sub-mm range. The receivers at sub-mm
wavelengths are of the heterodyne type, which means that the signal
can be fixed at a high frequency stability ($\sim 10^{-12}$).
Besides, FIR lines can be observed at early cosmological epochs ($z
\gtrsim 10$) which are far beyond the range accessible to optical
observations ($z\lesssim 4$).

\section{Astronomically observed FS transitions}\label{SecAstro}

The ground state FS transitions in mid- and far-infrared are
observed in emission in the interstellar dense and cold molecular
gas clouds, diffuse ionized gas in the star forming  H~\textsc{ii}
regions and in the `coronal' gas of active galactic nuclei (AGNs),
and in the warm gas envelopes of the protostellar objects. Cold
molecular gas clouds have been observed not only in our Galaxy, but
also in numerous galaxies with redshifts $z > 1$ up to $z = 6.42$
\cite{MCC05} and often around powerful quasars and radio galaxies
\cite{Omo07}. Recently the C~\textsc{ii} 158 $\mu$m line and CO low
rotational lines were used to set a limit on the variation of the
product $\mu\alpha^2$ at $z = 4.69$ and 6.42 \cite{LRK07}. The FIR
transitions in C\,{\sc i} (370, 609 $\mu$m) were detected at
$z=2.557$ towards H1413+117 \cite{WHD03,WDH05}. Four other
observations of the C\,{\sc i} 609 $\mu$m line were reported at $z =
4.120$ (PSS 2322+1944) \cite{PBC04}, at $z = 2.285$ (IRAS
F10214+4724) and $z=2.565$ (SMM J14011+0252) \cite{WDH05}, and at $z
= 3.913$ (APM 08279+5255) \cite{WWN06}.

In our Galaxy the most luminous protostellar objects are seen in the
O~\textsc{i} lines $\lambda\lambda63, 146$ $\mu$m\, \cite{CHT96} and
in the FIR lines from intermediate ionized atoms O~\textsc{iii},
N~\textsc{iii}, N~\textsc{ii} and C~\textsc{ii}, photoionized by the
stellar continuum \cite{BP03}. The lines of N~\textsc{ii} (122, 205
$\mu$m) S~\textsc{iii} (19, 34 $\mu$m), Fe~\textsc{iii} (23 $\mu$m),
Si~\textsc{ii} (35 $\mu$m), Ne~\textsc{iii} (36 $\mu$m),
O~\textsc{iii} (52, 88 $\mu$m), N~\textsc{iii} (57 $\mu$m),
O~\textsc{i} (63, 146 $\mu$m), and C~\textsc{ii} (158 $\mu$m)
\cite{GDG77, MSF80, CHE93}, as well as Ne~\textsc{ii} (13 $\mu$m),
S~\textsc{iv} (11 $\mu$m), and Ar~\textsc{iii} (9 $\mu$m)
\cite{GWD81} have been observed in the highly obscured ($A_v \simeq
21$ mag) massive star forming region G333.6--0.2. The FS transitions
of N~\textsc{iii}, O~\textsc{iii}, Ne~\textsc{iii}, S~\textsc{iii},
Si~\textsc{ii}, N~\textsc{iii}, O~\textsc{i}, C~\textsc{ii}, and
N~\textsc{ii} are detected in numerous Galactic H~\textsc{ii}
regions \cite{SCR95, SCC97, SRC04}. Compact and ultracompact
H~\textsc{ii} regions are the sources of the FS lines of
S~\textsc{iii}, O~\textsc{iii}, N~\textsc{iii}, Ne~\textsc{ii},
Ar~\textsc{iii}, and S~\textsc{iv}\, \cite{ACW97, OKY03}. Giant
molecular clouds in the Orion Kleinmann-Low cluster \cite{LBS06},
the Sgr~B2 complex \cite{GRC03, PBS07}, the $\rho$~Oph and
$\sigma$~Sco star-forming regions \cite{OON06}, and in the Carina
nebular \cite{MOS04, OPS06} emit the FIR lines of O~\textsc{i},
N~\textsc{ii}, C~\textsc{ii}, Si~\textsc{ii}, O~\textsc{iii}, and
N~\textsc{iii}.

Ions with low excitation potential $E_{\rm ex} < 50$ eV
(N~\textsc{ii}, Fe~\textsc{ii}, S~\textsc{iii}, Ar~\textsc{iii},
Fe~\textsc{iii}) as well as ions with high excitation potential 50
eV $ < E_{\rm ex} \leq 351$ eV (O~\textsc{iii}, Ne~\textsc{iii},
Ne~\textsc{v}, Mg~\textsc{v}, Ca~\textsc{v}, Na~\textsc{vi},
Mg~\textsc{vii}, Si~\textsc{vii}, Ca~\textsc{vii}, Fe~\textsc{vii},
Si~\textsc{ix}) are effectively produced by hard ionizing radiation
and ionzing shocks in the gas surrounding active galactic nuclei.
The FS emission lines of these ions have been detected with the {\it
Infrared Space Observatory (ISO)} and the {\it Spitzer Space
Telescope (Spitzer)} in Seyfert galaxies, 3C radio sources and
quasars, and in ultraluminous infrared galaxies in the redshift
interval from $z \sim 0.01$ up to $z = 0.994$
\cite{GC00,SLV02,DSA06,DWS07,SVH07,GCWL07}.

The infrared FS lines of the neutral atoms Si~\textsc{i},
S~\textsc{i}, and Fe~\textsc{i} have not been detected yet in
astronomical objects, but these atoms were observed in resonance
ultraviolet lines in two damped Ly$\alpha$ systems at $z = 0.452$
\cite{VD07} and $z = 1.15$ \cite{QRB08} toward the quasars HE
0000--2340 and HE 0515--4414, respectively.

The FIR lines are expected to be observed in extragalactic objects
at a new generation of telescopes such as the Stratospheric
Observatory for Infrared Astronomy (SOFIA), the Herschel Space
Observatory originally called `FIRST' for `Far InfraRed and
Submillimeter Telescope', and the Atacama Large Millimeter Array
(ALMA) which open a new opportunity of probing the relative values
of the fundamental physical constants with an extremely high
accuracy ($\delta \sim 10^{-7}$) locally and at different
cosmological epochs.

\section{Estimate of the sensitivity coefficients}
\label{estimate}

In the nonrelativistic limit and for an infinitely heavy point-like
nucleus all atomic transition frequencies are proportional to the
Rydberg constant, $\mathcal{R}$. In this approximation, the ratio of
any two atomic frequencies does not depend on any fundamental
constants. Relativistic effects cause corrections to atomic energy,
which can be expanded in powers of $\alpha^2$ and $(\alpha Z)^2$,
the leading term being $(\alpha Z)^2\mathcal{R}$, where $Z$ is
atomic number. Corrections accounting for the finite nuclear mass
are proportional to $\mathcal{R}/(\mu Z)$, but for atoms they are
much smaller than relativistic corrections. The finite nuclear mass
effects form the basis for the molecular constraints to the $m_{\rm
p}/m_{\rm e}$ mass ratio variation
\cite{RBH06,FK07a,T75,P77,VL93,PIV98,MSIV06}.

Consider the dependence of an atomic frequency $\omega$ on $\alpha$
in the co-moving reference frame:
 \begin{align}\label{qfactor1}
 \omega_z = \omega + q x + \dots, \quad x \equiv
 \left({\alpha_z}/{\alpha}\right)^2 - 1\, .
 \end{align}
Here $\omega$ and $\omega_z$ are the frequencies corresponding to
the present-day value of $\alpha$ and to a change $\alpha
\rightarrow \alpha_z$ at a redshift $z$. The parameter $q$
(so-called $q$-factor) is individual for each atomic transition
\cite{DFK02}.

If $\alpha$ is not a constant, the parameter $x$ differs from zero
and the corresponding frequency shift, $\Delta\omega = \omega_z -
\omega$, is given by:
 \begin{align}\label{qfactor2}
 {\Delta\omega}/{\omega} = 2\mathcal{Q}\,({\Delta\alpha}/{\alpha})\,,
 \end{align}
where ${\cal Q} = q/\omega$ is the dimensionless sensitivity
coefficient, and $\Delta\alpha/\alpha \equiv (\alpha_z -
\alpha)/\alpha$. Here we assume that $|\Delta\alpha/\alpha| \ll 1$.

If such a frequency shift takes place for a distant object observed
at a redshift $z$, then an apparent change in the redshift,
$\Delta z = \tilde{z} - z$, occurs:
 \begin{align}\label{qfactor3}
 {\Delta\omega}/{\omega} = -\Delta z/(1+z) \equiv {\Delta v}/{c}\, ,
 \end{align}
where $\Delta v$ is the Doppler radial velocity shift. If $\omega'$
is the observed frequency from a distant object, then the true
redshift is given by
 \begin{align}\label{zfactor1}
 1+z = \omega_z/\omega' \, ,
 \end{align}
whereas the shifted (apparent) value is
 \begin{align}\label{zfactor2}
 1+\tilde{z} = \omega/\omega' \, .
 \end{align}
If we have two lines of the same element with the apparent redshifts
$\tilde{z}_1$ and $\tilde{z}_2$ and the corresponding sensitivity
coefficients ${\cal Q}_1$ and ${\cal Q}_2$, then
 \begin{align}\label{zfactor3}
 2\Delta {\cal Q}(\Delta\alpha/\alpha) = (\tilde{z}_1 -
 \tilde{z}_2)/(1 + z ) = \Delta v /c\, ,
 \end{align}
where $\Delta v = v_1 - v_2$ is the difference of the measured
radial velocities of these lines, and
$\Delta {\cal Q} = {\cal Q}_2 - {\cal Q}_1$.

Relativistic corrections grow with atomic number $Z$, but for
optical and UV transitions in light atoms they are small, i.e.
$\mathcal{Q} \sim (\alpha Z)^2\ll 1$. For example, Fe~\textsc{ii}
lines have sensitivities $\mathcal{Q} \sim 0.03$ \citep{PKT07}.
Other atomic transitions, used in astrophysical searches for
$\alpha$-variation have even smaller sensitivities. The only
exceptions are the Zn~\textsc{ii} $\lambda 2026$ \AA\ line, where
$\mathcal{Q} \approx 0.050$ \citep{DFK02} and the Fe~\textsc{i}
resonance transitions considered in \citep{DF08} where $\mathcal{Q}$
ranges between 0.03 and 0.09. One can significantly increase the
sensitivity to $\alpha$-variation by using transitions between FS
levels of one multiplet \cite{DF05b}. In the nonrelativistic limit
$\alpha \rightarrow 0$ such levels are exactly degenerate.
Corresponding transition frequencies $\omega$ are approximately
proportional to $(\alpha Z)^2$. Consequently, for these transitions
$\mathcal{Q}\approx 1$ and
 \begin{align}\label{qfactor4}
 {\Delta\omega}/{\omega}
 \approx
 2 {\Delta\alpha}/{\alpha}\, ,
 \end{align}
which implies that for any two FS transitions $\Delta {\cal Q}
\approx 0$. In this approximation \daa\ cannot be determined from
\Eref{zfactor3}.

We will show now that in the next order in $(\alpha Z)^2$ the
$\mathcal{Q}$-factors of the FS transitions deviate from unity and
$\Delta \mathcal{Q}$ in \Eref{zfactor3} is not equal to zero. In
fact, for heavy atoms with $\alpha Z \sim 1$ it is possible to find
FS transitions with $|\Delta \mathcal{Q}| \gg 1$ \cite{DF05b}. Here
we focus on atoms with $\alpha Z \ll 1$, which are more important
for astronomical observations. For such atoms
 $|\Delta \mathcal{Q}| < 1$ and, as we will show below, there is a
simple analytical relation between $\Delta \mathcal{Q}$ and
experimentally observed FS intervals.

There are two types of relativistic corrections to atomic energy.
The first type depends on the powers of $\alpha Z$ and rapidly grows
along the periodic table. The second type of corrections depends on
$\alpha$ and does not change much from atom to atom. Such
corrections are usually negligible, except for the lightest atoms.
Expanding the energy of a level of the FS multiplet $^{2S+1}\!L_J$
into (even) powers of $\alpha Z$ we have (see \cite{Sob79},
Sec.~5.5):
 \begin{align}\nonumber
 E_{L,S,J} &= E_0
 +\tfrac{A(\alpha Z)^2}{2}\left[J(J+1)-L(L+1)-S(S+1)\right]
 \nonumber\\
 &+B_J\,(\alpha Z)^4 + \dots\,,
 \label{FS1}
 \end{align}
where $A$ and $B_J$ are the parameters of the FS multiplet. Note,
that in general, $B_J$ depends on quantum numbers $L$ and $S$, but
we will omit $L$ and $S$ subscripts since they do not change the
following discussion. In \Eref{FS1} we keep the term of the
expansion $\sim(\alpha Z)^4$, but neglect the term $\sim\alpha^2$.
This is justified only for atoms with $Z\gtrsim 10$. Therefore, the
following discussion is not applicable to atoms of the second
period. As long as these atoms are very important for astrophysics,
we will briefly discuss them in the end of this section.

The strongest FS transitions are of {\it M1}-type. They occur between
levels with $\Delta J=1$:
 \begin{align}\label{FS2}
 \omega_{J,J-1}
 &=E_{L,S,J}-E_{L,S,J-1}
 \nonumber\\
 &=AJ(\alpha Z)^2+\left(B_J-B_{J-1}\right)(\alpha Z)^4\, .
 \end{align}
In the first order in $(\alpha Z)^2$ we have the well known
Land\'{e} rule: $\omega_{J,J-1} = AJ(\alpha Z)^2$, which directly
leads to \Eref{qfactor4}. In the next order we get:
 \begin{align}\label{FS3}
 \mathcal{Q}_{J,J-1} = 1 + \frac{B_J-B_{J-1}}{AJ}\,(\alpha Z)^2 \,.
 \end{align}

Let us consider the multiplet $^3\!P_J$ (i.e. the ground multiplet
for Si~\textsc{i}, S~\textsc{i}, Ar~\textsc{iii}, Mg~\textsc{v},
Ca~\textsc{v}, Na~\textsc{vi}, Mg~\textsc{vii}, Si~\textsc{vii},
Ca~\textsc{vii}, and Si~\textsc{ix}). For two transitions
$\omega_{2,1}$ and $\omega_{1,0}$ \Eref{FS3} gives:
 \begin{align}\label{FS4}
 \mathcal{Q}_{2,1}-\mathcal{Q}_{1,0} = \frac{B_2-3B_1+2B_0}{2A}\,(\alpha Z)^2\,.
 \end{align}
At the same time, \Eref{FS2} gives the following expression for the
frequency ratio:
 \begin{align}\label{FS5}
 \frac{\omega_{2,1}}{\omega_{1,0}}
 &= 2 + \frac{B_2-3B_1+2B_0}{A}\,(\alpha Z)^2\,.
 \end{align}
Comparison of Eqs.~\eqref{FS4} and  \eqref{FS5} leads to the final result:
 \begin{align}\label{FS6}
 \Delta \mathcal{Q} =  \mathcal{Q}_{2,1}-\mathcal{Q}_{1,0}
 = \frac{1}{2}\,\left(\frac{\omega_{2,1}}{\omega_{1,0}}\right)-1\,.
 \end{align}
In a general case of the $^{2S+1}\!L_J$ multiplet the difference
between the sensitivity coefficients $\mathcal{Q}_{J,J-1}$ and
$\mathcal{Q}_{J-1,J-2}$ is given by
 \begin{align}\label{FS7}
 \Delta \mathcal{Q} = \frac{J-1}{J}\,
 \left(\frac{\omega_{J,J-1}}{\omega_{J-1,J-2}}\right)-1\,.
 \end{align}
If two arbitrary FS transitions $\omega_{J_1,J_1'}$ and $\omega_{J_2,J_2'}$
of the $^{2S+1}\!L_J$ multiplet are considered, then
the difference
 $\Delta\mathcal{Q}=\mathcal{Q}_{J_2,J_2'}-\mathcal{Q}_{J_1,J_1'}$
is expressed by
 \begin{align}\label{FS7a}
 \Delta \mathcal{Q}
 = \frac{J_1(J_1+1) - J_1'(J_1'+1)}{J_2(J_2+1) - J_2'(J_2'+1)}\,
 \left(\frac{\omega_{J_2,J_2'}}{\omega_{J_1,J_1'}}\right) - 1\,.
 \end{align}
This equation can be used also for $E2$-transitions with $\Delta
J=2$ and for combination of $M1$- and $E2$-transitions.

It is to note that the derived values of $\Delta \mathcal{Q}$ for
two FS transitions are expressed in terms of their frequencies,
which are known from the laboratory measurements. Another point is
that the right-hand side of \Eref{FS7} turns to zero when the
frequency ratio equals $J/(J-1)$, i.e. when the Land\'{e} rule is
fulfilled. Eqs.~\eqref{FS6} --~\eqref{FS7a} hold only as long as we
neglect corrections of the order of $\alpha^2$ and $(\alpha Z)^6$ to
\Eref{FS1}, which is justified for the atoms in the middle of the
periodic table, i.e. approximately from Na ($Z=11$) to Sn ($Z=50$).


\begin{table*}[tbh]

\caption{The differences of the sensitivity coefficients $\Delta
\mathcal{Q}$ of the FS emission lines within the ground multiplets
$^3\!P_J$, $^5\!D_J$, $^6\!D_J$, $^4\!F_J$, and $^3\!F_J$ for the
most abundant atoms and ions. The FS intervals for S~\textsc{i},
Fe~\textsc{i--iii}, Ar~\textsc{iii},  Mg~\textsc{v}, Ca~\textsc{v},
and Si~\textsc{vii} are inverted. The excitation temperature $T_{\rm
ex}$ for the upper level is indicated. Transition wavelengths and
frequencies (rounded) are taken from Ref.~\cite{NIST}. The values of
$\Delta \mathcal{Q}$ for the ions C~\textsc{i}, N~\textsc{ii}, and
O~\textsc{iii} are calculated using Eq.~(5.197) from
Ref.~\cite{Sob79}.} \label{tab1}
\begin{tabular}{lcdddcdddcd}
\hline\hline\\[-7pt]
 \multicolumn{1}{c}{Atom/Ion}
 &\multicolumn{4}{c}{Transition $a$}
 &\multicolumn{4}{c}{Transition $b$}
 &\multicolumn{1}{c}{$\omega_b/\omega_a$}
 &\multicolumn{1}{c}{$\Delta \mathcal{Q}=$} \\
 &\multicolumn{1}{c}{$(J_a,J_a')$}
 &\multicolumn{1}{c}{$\lambda_a$ ($\mu$m)}
 &\multicolumn{1}{c}{$\omega_a$ (cm$^{-1}$)}
 &\multicolumn{1}{c}{$T_{\rm ex}$ (K)}
 &\multicolumn{1}{c}{$(J_b,J_b')$}
 &\multicolumn{1}{c}{$\lambda_b$ ($\mu$m)}
 &\multicolumn{1}{c}{$\omega_b$ (cm$^{-1}$)}
 &\multicolumn{1}{c}{$T_{\rm ex}$ (K)}
&&\multicolumn{1}{c}{$\mathcal{Q}_b-\mathcal{Q}_a$} \\
  \hline\\[-5pt]
C~\textsc{i}   &     (1,0) &609.1& 16.40&  24&  (2,1)     &370.4&  27.00&  63&1.646& -0.008 \\
Si~\textsc{i}  &     (1,0) &129.7& 77.11& 111&  (2,1)     & 68.5& 146.05& 321&1.894& -0.053 \\
S~\textsc{i}   &     (0,1) & 56.3&177.59& 825&  (1,2)     & 25.3& 396.06& 570&2.230&  0.115 \\
Ti~\textsc{i}  &     (2,3) & 58.8&170.13& 245&  (3,4)     & 46.1& 216.74& 557&1.274& -0.045 \\
Fe~\textsc{i}  &     (2,3) & 34.7&288.07&1013&  (3,4)     & 24.0& 415.93& 599&1.444&  0.083 \\
               &     (1,2) & 54.3&184.13&1278&  (2,3)     & 34.7& 288.07&1013&1.565&  0.043 \\
               &     (0,1) &111.2& 89.94&1407&  (1,2)     & 54.3& 184.13&1278&2.048&  0.024 \\
N~\textsc{ii}  &     (1,0) &205.3& 48.70&  70&  (2,1)     &121.8&  82.10& 188&1.686& -0.016 \\
Fe~\textsc{ii} & (5/2,7/2) & 35.3&282.89& 961&  (7/2,9/2) & 26.0& 384.79& 554&1.360&  0.058 \\
               & (3/2,5/2) & 51.3&194.93&1241&  (5/2,7/2) & 35.3& 282.89& 961&1.451&  0.037 \\
               & (1/2,3/2) & 87.4&114.44&1406&  (3/2,5/2) & 51.3& 194.93&1241&1.703&  0.022 \\
O~\textsc{iii} &     (1,0) & 88.4&113.18& 163&  (2,1)     & 51.8& 193.00& 441&1.705& -0.027 \\
S~\textsc{iii} &     (1,0) & 33.5&298.69& 430&  (2,1)     & 18.7& 534.39&1199&1.789& -0.105 \\
Ar~\textsc{iii}&     (0,1) & 21.9&458.05&2259&  (1,2)     &  9.0&1112.18&1600&2.428&  0.214 \\
Fe~\textsc{iii}&     (2,3) & 33.0& 302.7&1063&  (3,4)     & 22.9& 436.2 & 628&1.441&  0.081 \\
               &     (1,2) & 51.7& 193.5&1342&  (2,3)     & 33.0& 302.7 &1063&1.564&  0.043 \\
               &     (0,1) &105.4& 94.9 &1478&  (1,2)     & 51.7& 193.5 &1342&2.039&  0.019 \\
Mg~\textsc{v}  &     (0,1) & 13.5& 738.7&3628&  (1,2)     &  5.6&1783.1 &2566&2.414&  0.207 \\
Ca~\textsc{v}  &     (0,1) & 11.5& 870.9&4713&  (1,2)     &  4.2&2404.7 &3460&2.761&  0.381 \\
Na~\textsc{vi} &     (1,0) & 14.3& 698  &1004&  (2,1)     &  8.6& 1161  &2675&1.663& -0.168 \\
Fe~\textsc{vi} & (5/2,3/2) & 19.6& 511.3& 736&  (7/2,5/2) & 14.8& 677.0 &1710&1.324& -0.054 \\
               & (7/2,5/2) & 14.8& 677.0&1710&  (9/2,7/2) & 12.3& 812.3 &2879&1.200& -0.067 \\
Mg~\textsc{vii}&     (1,0) & 9.0 & 1107 &1593&  (2,1)     &  5.5& 1817  &4207&1.641& -0.179 \\
Si~\textsc{vii}&     (0,1) & 6.5 & 1535 &8007&  (1,2)     &  2.5& 4030  &5817&2.625&  0.313 \\
Ca~\textsc{vii}&     (1,0) & 6.2 &1624.9&2338&  (2,1)     &  4.1&2446.5 &5858&1.506& -0.247 \\
Fe~\textsc{vii}&     (3,2) & 9.5 &1051.5&1513&  (4,3)     &  7.8&1280.0 &3354&1.217& -0.087 \\
Si~\textsc{ix} &     (1,0) &  3.9&2545.0&3662&  (2,1)     &  2.6&3869 &9229&1.520& -0.240 \\
  \hline\hline
\end{tabular}
\end{table*}

\tref{tab1} lists the calculated $\Delta \mathcal{Q}$ values for the
most abundant atoms and ions observed in Galactic and extragalactic
gas clouds. The ions C~\textsc{i}, Si~\textsc{i}, N~\textsc{ii},
O~\textsc{iii}, Na~\textsc{vi}, Mg~\textsc{vii}, and
Ca~\textsc{vii}, have configuration $ns^2 np^2$ and `normal' order
of the FS sub-levels. The ions Mg~\textsc{v}, Si~\textsc{vii},
S~\textsc{i}, and Ca~\textsc{v} have configuration $ns^2 np^4$ and
`inverted' order of the FS sub-levels. However, \Eref{FS6} is
applicable for both cases. We note that the FS lines of
N~\textsc{ii} (122, 205 $\mu$m) can be asymmetric and broadened due
to hyperfine components, as observed in \cite{LBS06,PBS07}. The
hyperfine splitting occurs also in the FS lines of Na~\textsc{vi}
(8.6, 14.3 $\mu$m).

Transition wavelengths and frequencies listed in \tref{tab1} are
approximate and are given only to identify the FS transitions.
At present, many of them have been measured with a sufficiently
high accuracy \cite{NIST}.

The Iron ions Fe~\textsc{i}, Fe~\textsc{ii}, Fe~\textsc{iii},
Fe~\textsc{vi}, and Fe~\textsc{vii}, have ground multiplets $^5\!D$,
$^6\!D$, $^5\!D$, $^4\!F$, and $^3\!F$, respectively. All these
multiplets, except the last one, produce more than two FS lines,
which can be used to further reduce the systematic errors. The
sensitivity coefficients for transitions in Iron and Titanium from
\tref{tab1} are calculated with the help of \Eref{FS7}.

According to \tref{tab1}, the absolute values of the difference
$\Delta\mathcal{Q}$ are usually quite large even for atoms with
$Z\sim 10$. The sign of $\Delta\mathcal{Q}$ is negative for atoms
with configuration $ns^2 np^2$ and positive for atoms with
configuration $ns^2 np^4$. These features are not surprising if we
consider the level structure of the respective configurations
\cite{Sob79}. Both of them have three terms: $^3\!P_{0,1,2}$,
$^1\!D_2$, and $^1\!S_0$, but for the configuration $ns^2 np^4$, the
multiplet $^3\!P_J$ is `inverted'. The splitting between these terms
is caused by the residual Coulomb interaction of $p$-electrons and
is rather small compared to the atomic energy unit $2\mathcal{R}$.

For example, the level $^1\!D_2$ for Si~\textsc{i} lies only
6299~cm$^{-1}$ above the ground state, which corresponds to
$E_D-E_P=0.029$~a.u.. Relativistic corrections to the energy are
dominated by the spin-orbit interaction, which for $p$-electrons has
the order of $0.1(\alpha Z)^2$~a.u.. The diagonal part of this
interaction leads to the second term in \Eref{FS1}, i.e. $A\,(\alpha
Z)^2\sim 200$~cm$^{-1}$. In the second order the non-diagonal
spin-orbit interaction causes repulsion between the levels $^3\!P_2$
and $^1\!D_2$ and results in non-zero parameter $B_2$. We can
estimate this correction as $B_2(\alpha Z)^4\sim A^2(\alpha
Z)^4/(E_P-E_D)\,\sim -10~\mathrm{cm}^{-1}$. This estimate has an
expected order of magnitude. Note that $B_2$ is negative. For normal
multiplets it reduces the ratio $\omega_{2,1}/\omega_{1,0}$, whereas
for the inverted multiplet the ratio increases. We see that this is
in a qualitative agreement with \tref{tab1}. Iron and Titanium ions
have configurations $3d^k 2s^l$, with $k=6,\,l=2$ for Fe~\textsc{i}
and $k=2,\,l=0,2$ for Fe~\textsc{vii} and Ti~\textsc{i}
respectively. As we can see from \tref{tab1}, here also all normal
multiplets (for Ti~\textsc{i}, Fe~\textsc{vi}, and Fe~\textsc{vii})
have negative values of $\Delta\mathcal{Q}$, while inverted
multiplets for all other ions have positive values of
$\Delta\mathcal{Q}$.

Equation~(\ref{FS4}) shows that sensitivity to $\alpha$-variation
grows with $Z$. For heavy atoms, $\alpha Z \sim 1$, neglected terms
in expansion \eqref{FS1} become important. That breaks relation
\eqref{FS7} between $\Delta \mathcal{Q}$ and FS intervals and
sensitivity coefficients $\mathcal{Q}$ have to be calculated
numerically. According to \tref{tab1}, the largest coefficients
$B_J$ appear for Ca~\textsc{v} and Si~\textsc{vii}. The neglected
corrections to $\Delta \mathcal{Q}$ can be estimated as $\sim
[A(\alpha Z)^2/(E_P-E_D)]^2 $, i.e. the uncertainty in $\Delta
\mathcal{Q}$ for Ca~\textsc{v} and Si~\textsc{vii} is less than
20\%. For other elements listed in \tref{tab1} this correction
should be smaller. Note that for Iron ions, which have the largest
$Z$, the relativistic effects are suppressed, because for
$d$-electrons they are typically an order of magnitude smaller, than
for $p$-electrons.

For light elements the accuracy of our estimate depends on the
neglected terms $\sim \alpha^2$. The discussion of these terms can
be found in \cite{Sob79} (see Eq.~(5.197) and Table~5.21 therein).
The corresponding correction decreases from almost 50\% for
Na~\textsc{vi} to 30\% for Mg~\textsc{vii} and to 15\% for
Si~\textsc{ix}.

For atoms with $Z\lesssim 10$ one can calculate $\Delta \mathcal{Q}$
using Eq.~(5.197) from Ref.~\cite{Sob79}. For example, for
C~\textsc{i}, N~\textsc{ii}, and O~\textsc{iii}, we get $\Delta
\mathcal{Q} = -0.008$, $-0.016$, and $-0.027$, respectively. As
expected, these values are much smaller than those for the heavier
elements. On the other hand, these ions are so important for
astrophysics, that we keep them in \tref{tab1}.

Numerical calculations for heavy many-electron atoms
are rather difficult to perform and the computed $\Delta
\mathcal{Q}$ values may not be very accurate. For atoms with $\alpha
Z \ll 1$ one can use \Eref{FS7} to check the accuracy of the
numerical results.

\begin{table}[bh]
  \caption{The differences between sensitivity coefficients of the FS transitions
within ground $^6\!D_J$ multiplet of Fe~\textsc{ii},
$\Delta \mathcal{Q} \equiv \mathcal{Q}_{J,J-1}-\mathcal{Q}_{J-1,J-2}$.
In the third column
we use calculated $q$-factors from \cite{PKT07} (see Table~I from this Ref.,
basis set [7$spdf$]).  In the fourth and fifth columns we apply \Eref{FS7} to
calculated and experimental FS intervals, respectively.}
  \label{tab2}
  \begin{tabular}{ccddd}
  \hline\hline
  \multicolumn{2}{c}{Transitions}
                        &\multicolumn{3}{c}{$\Delta\mathcal{Q}$}\\
 (5/2,7/2) &  (7/2,9/2) & 0.045 & 0.049  & 0.058 \\
 (3/2,5/2) &  (5/2,7/2) & 0.023 & 0.029  & 0.037 \\
 (1/2,3/2) &  (3/2,5/2) & 0.017 & 0.016  & 0.022 \\
  \hline\hline
  \end{tabular}
\end{table}

As an example we consider the ground $^6\!D_J$ multiplet of
Fe~\textsc{ii} ion (\tref{tab2}). One can see that numerical results
in Ref.~\cite{PKT07} are in good agreement with the values obtained
from \Eref{FS7} for the calculated FS intervals. However, when we
apply \Eref{FS7} to actual experimental FS intervals, the agreement
worsens noticeably. It is well known that deviations from the
Land\'{e} rule for FS intervals depend on the interplay between the
(non-diagonal) spin-orbit and the residual Coulomb interactions
\cite{Sob79}. For this reason numerical results are very sensitive
to the treatment of the effects of the core polarization and the
valence correlations. Note also that the calculated $q$-factors are
firstly used to find sensitivity coefficients $\mathcal{Q}$, and
then the (small) differences are taken. Obviously this makes the
whole calculation rather unstable. Similarly, \Eref{FS7} can be used
to check calculations of the $q$-factors for other atoms considered
in Refs.~\cite{BFK05a,BFK06,DF08}.

Numerical calculations for light atoms with $Z\lesssim 10$ are
usually much simpler and more reliable. However, as we have pointed
out above, the differences in the sensitivity coefficients of the
light atoms depend on the relativistic corrections $\sim\alpha^2$.
This means that the Breit interaction between valence electrons
should be accurately included, while the majority of the published
results were obtained in the Dirac-Coulomb approximation.

There is a certain similarity between the present method and the
method of optical doublets, used previously to study
$\alpha$-variation (see, e.g., \cite{BSS04} and references therein).
In that method, however, the FS energy constitutes a small fraction
of the total transition energy. Therefore, the parameter
$\Delta\mathcal{Q}$ for optical transitions is much smaller. Note
that for the mid- and far-infrared FS lines, the transition energy
and the FS splitting coincide, which leads to a much larger
parameter $\Delta\mathcal{Q}$.

\section{Discussion and Conclusions}

In this paper we suggest to use two, or more FS lines of the same
ion to study possible variation of $\alpha$ at early stages of the
evolution of the Universe up to $\Delta T \sim 96$\% of $T_{\rm U}$.
The sensitivity of the suggested method is proportional to
$\Delta\mathcal{Q}$, as seen from \Eref{zfactor3}. We have deduced a
simple analytical expression to calculate $\Delta\mathcal{Q}$ for
the FS transitions in light atoms and ions within the range of
nuclear charges $11 \le Z\le 26$. We found that
$|\Delta\mathcal{Q}|$ grows with $Z$ and reached 0.2~--~0.4 for the
ions of Ar and Ca. This is about one order of magnitude higher than
typical sensitivities in the optical and UV range.

In addition of being more sensitive, this method provides also a
considerable reduction of
the Doppler noise, which limits the accuracy of the optical
observations. Using the lines of the same element reduces the
sources of the Doppler noise to the inhomogeneity of the excitation
temperature $T_{\rm ex}$ within the cloud(s). Alternatively, when
the lines of different species are used, the Doppler noise may be
significantly higher because of the difference of the respective
spatial distributions.

At present, the precision of the existing radio observations of the
FS lines from distant objects is considerably lower than in the most
accurate optical observations. For example, the error in the line
center position for the C~\textsc{i} $J = 2 \rightarrow 1$ and $J =
1 \rightarrow 0$ lines at $z = 2.557$ was $\sigma_{v,{\rm radio}} =
8$ and 25 \kms\, respectively \cite{WHD03,WDH05}. This has to be
compared with the precision of the modern optical measurements of
$\sigma_{v,{\rm opt}} = 85$ \ms~\cite{LML07,LCM06}. In the optical
range the error $\sigma_{v,{\rm opt}}$ includes both random and
systematic contributions. The systematic error is the wavelength
calibration error which is negligible at radio frequencies.

In the forthcoming observations with ALMA, the statistical error is
expected to be several times smaller than 85 \ms. Together with the
higher sensitivity to $\alpha$-variation, this would allow to
estimate \daa\ at the level of one tenth of ppm~--- well beyond the
limits of the contemporary optical observations and comparable to
the anticipated sensitivity of the next generations of spectrographs
for the VLT and the EELT \cite{MML06,M07}. Thus, FIR lines offer a
very promising strategy to probe the hypothetical variability of the
fine-structure constant both locally and in distant extragalactic
objects.

\begin{acknowledgments}
MGK, SGP, and SAL gratefully acknowledge the hospitality of
Hamburger Sternwarte while visiting there. This research has been
partly supported by the DFG projects SFB 676 Teilprojekt C,
the RFBR grants No. 06-02-16489 and 07-02-00210, and by the Federal
Agency for Science and Innovations grant NSh 9879.2006.2.
\end{acknowledgments}


\end{document}